\documentclass[11pt,twoside]{article}
\usepackage{asp2010}

\resetcounters

\begin{document}

\title{Elemental and isotopic abundances and chemical evolution of galaxies}
\author{Chiaki Kobayashi$^1$}
\affil{$^1$Centre for Astrophysics Research, School of Physics, Astronomy and Mathematics, University of Hertfordshire}

\def\gtsim {>\kern-1.2em\lower1.1ex\hbox{$\sim$}~}   
\def\ltsim {<\kern-1.2em\lower1.1ex\hbox{$\sim$}~}   
\def \apj {ApJ}
\def \apjs {ApJS}
\def \aj  {AJ}
\def \aap {A\&A} 
\def \mnras {MNRAS}
\def \pasj {PASJ}
\def \araa {ARA\&A}
\def \pasa {PASA}

\begin{abstract}
Elemental and isotopic abundances are the fossils of galactic archaeology. The observed [X/Fe]-[Fe/H] relations in the Galactic bulge and disk and the mass-metallicity relation of galaxies are roughly reproduced with chemodynamical simulations of galaxies under the standard $\Lambda$-CDM picture and standard stellar physics.
The isotopic ratios such as $^{17,18}$O and $^{25,26}$Mg may require a refinement of modelling of supernova and asymptotic giant branch stars. The recent observation of the Carbon-rich damped Lyman $\alpha$ system can be reproduced only with faint core-collapse supernovae. This suggests that chemical enrichment by the first stars in the first galaxies is driven not by pair-instability supernovae but by core-collapse supernovae ($\sim 20-50M_\odot$). The observed F abundances can be reproduced with the neutrino processes of core-collapse supernovae. As in F, the observations of elemental abundances in small systems may requires further complications of chemical enrichment. In globular clusters the relative contribution from low-mass supernovae is likely to be smaller than in the field, while the contribution from massive supernovae seems smaller in dwarf spheroidal galaxies than in the solar neighborhood.
\end{abstract}

\vspace*{-10mm}
\section{Introduction}
\vspace*{-3mm}

Elemental and isotopic abundances are the fossils of galactic archaeology.
Different elements are produced from stars on different timescales, therefore
elemental and isotopic abundance ratios provide independent information on the ``age'' of a system and 
can be used as a form of ``cosmic clock''. 
The formation and evolutionary history of galaxies can be constrained in theoretical models 
by using the observed abundances of stars.  
The space astrometry missions (e.g., GAIA) and large-scale
surveys (e.g., HERMES on the AAT) will produce unprecedented information on the chemodynamical structure of the
Milky Way Galaxy. 
Theoretically \citet{kob06} succeeded in reproducing the average evolution of major elements 
(except for Ti) in the solar neighbourhood including normal Type II Supernovae (SNe II) and hypernovae (HNe).
In this paper we extend the discussion to isotope ratios (\S 2) and more realistic chemodynamical simulations (\S 3).
From the observed abundances, we discuss the possibility of non-standard chemical enrichment in the very metal-poor damped Lyman $\alpha$ system (DLA, \S 4), globular clusters (GCs, \S 5), and dwarf spheroidal galaxies (dSphs, \S 6).

\vspace*{-5mm}
\section{Isotope Ratios}
\vspace*{-3mm}

\begin{figure}[t]
\center
\includegraphics[width=9.5cm]{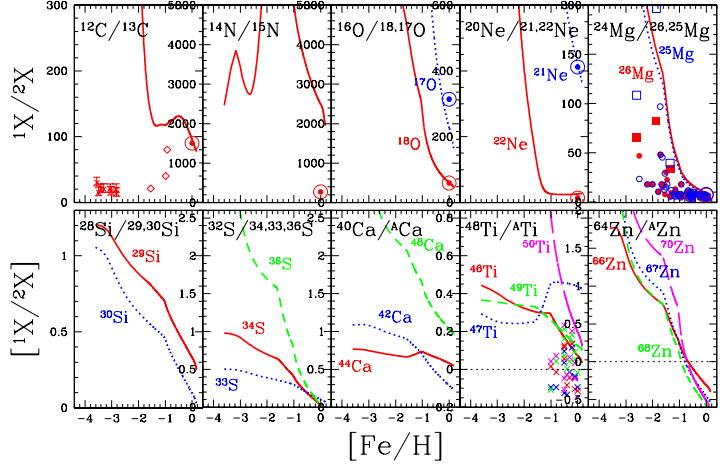}
\vspace*{-3mm}
\caption{\label{fig:iso}
Evolution of isotope ratios against [Fe/H]
for the solar neighbourhood 
with AGB yields.
See \citet{kob11agb} for the observational data sources for C, Mg, and Ti (dots).
The solar ratios are shown with the solar symbols at [Fe/H] $=0$ in the upper panels.
\vspace*{-5mm}
}
\end{figure}

We include the latest chemical enrichment input into our chemical evolution models, which are summarised in \citet{kob11agb}.
The basic equations of galactic chemical evolution are described in \citet{kob06}.
Figure \ref{fig:iso} shows the 
evolution of the isotope ratios against [Fe/H]
for the solar neighbourhood. 
In general, core-collapse supernovae are the main producers of the major isotopes with more 
minor isotopes synthesized at higher metallicity.
For this reason, the evolution of major to minor isotope ratios continuously decreases toward higher metallicity.
The slope changes at [Fe/H] $\sim -2.5$ and $\sim -1.5$ are due to the onset of  intermediate- and 
low-mass asymptotic giant branch (AGB) stars, respectively.
The rapid change in the slope at [Fe/H] $\sim -1$ is caused by Type Ia Supernovae (SNe Ia).

The time evolution of isotope ratios depends on the star formation and chemical enrichment histories of the system.
See \citet{kob11agb} for the models of bulge, thick disk, and halo.
In general, the ratios between the major and minor isotopes such as $^{24}$Mg/$^{25,26}$Mg 
are smaller in the bulge and thick disk, and are larger in the halo because of the metallicity 
effect of supernovae. However, the $^{16}$O/$^{17}$O ratio in the halo is low due to the production of 
$^{17}$O from low-mass AGB stars, as also seen in the high [(C, F)/Fe] abundances.
Therefore, the isotopic ratios can be used as a tool to pick out the stars that form in a 
system with a low chemical enrichment efficiency.
This may be possible in our halo, but more likely in small satellite galaxies that were accreted 
onto our Milky Way Galaxy.

\vspace*{-5mm}
\section{Chemodynamical Simulations}
\vspace*{-3mm}

\begin{figure}[t]
\center
\includegraphics[width=5.5cm,angle=-90]{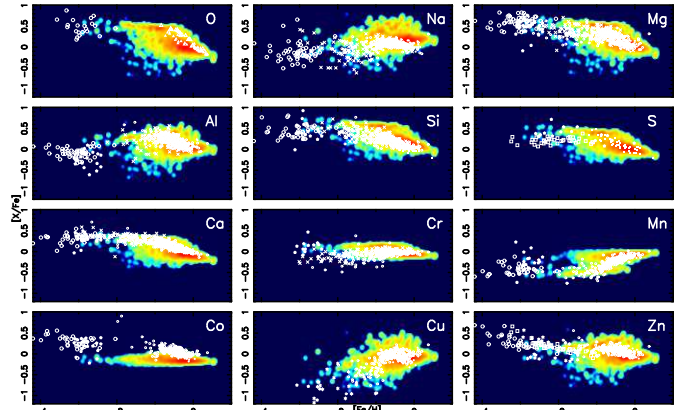}
\vspace*{-3mm}
\caption{\label{fig:xfe}
[X/Fe]-[Fe/H] relations in the solar neighborhood at $z=0$.
The contours show the frequency distribution of stars in the simulated galaxies, where red is for the highest frequency.
See \citet{kob11mw} for the observational data sources.
}
\vspace*{-5mm}
\end{figure}

In a real galaxy, the star formation history is not so simple and
the interstellar medium (ISM) is not homogeneous. In chemodynamical simulation \citep{kob11mw}, hydrodynamics and chemical enrichment are solved self-consistently through the galaxy formation.
Star formation and chemical enrichment depend on the local density and mixing due to dynamical effects such as merging and migration are also naturally included.
Figure \ref{fig:xfe} shows the frequency distribution of stars in the solar neighborhood at present. See \citet{kob11mw} for the bulge and the thick disk.

In our simulations, the metallicity of the first enriched stars reaches [Fe/H] $\sim -3$.
At later times, the star forming region becomes denser, and both metal richer and poorer stars than [Fe/H] $\sim -3$ appear.
Different from one-zone chemical evolution models, the following phenomena occur in the case of inhomogeneous enrichment:
i) The age-metallicity relation is weak. In other words, the most metal-poor stars are not always the oldest stars.
ii) SNe Ia can affect the elemental abundance ratios at [Fe/H] $\ltsim -1$ even with the metallicity inhibition of SNe Ia. The SN Ia contribution is characterised by low [$\alpha$/Fe] and high [Mn/Fe].
iii) The scatter of elemental abundance ratios becomes large if the supernova yield depends on progenitor metallicity such as Na and
iv) Some of CEMP/NEMP stars can be explained with the local enrichment from AGB stars even without the binary effect. In fact, the observed [N/O]-[O/H] trend can be reproduced with our simulation without including the effect of rotating massive stars.

\vspace*{-5mm}
\section{Carbon-Enhanced Dampled Lyman $\alpha$ (DLA) System}
\vspace*{-3mm}

\begin{figure}
\center
\includegraphics[width=7.5cm]{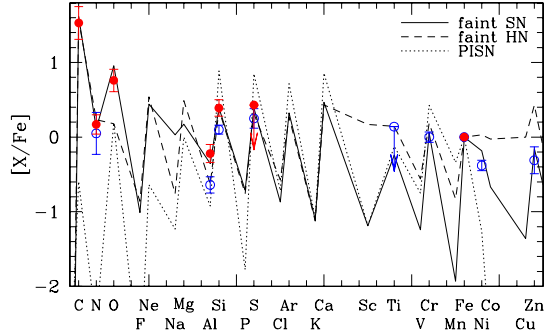}
\vspace*{-3mm}
\caption{\label{fig:dla}
The elemental abundance pattern of the metal-poor C-rich DLA (filled circles) and peculiar DLA (open circles).
The solid and short-dashed lines show the nucleosynthesis yields of faint core-collapse supernovae from $25M_\odot$ stars with mixing-fallback.
The dotted line is for PISNe from $170 M_\odot$ stars.
}
\vspace*{-5mm}
\end{figure}

The observations of very metal-poor DLAs have opened a new window to study the chemical enrichment of the Universe by the first generations of stars.
Figure \ref{fig:dla} shows the elemental abundance ratios from C to Zn relative to Fe.
We implement two cases with different explosion energies: $1 \times 10^{51}$ erg (SN, solid line) and $20 \times 10^{51}$ erg (HN, short-dashed line) for an initial mass of $M=25M_\odot$ and metallicity $Z=0$ (see \citealt{kob11dla} for the details).
An efficient mixing-fallback is adopted in both models, black hole masses are as large as $\sim 6 M_\odot$, and thus the ejected iron mass is much smaller than ``normal'' SNe/HNe in \citet{kob06} that are responsible for the chemical enrichment in the Galaxy.
Since C is synthesized in the outermost region of the ejecta, [C/Fe] ratio is higher for faint SNe/HNe than ``normal'' SNe/HNe.

We then compare these model calculations with the available abundance measurements 
in the extremely metal-poor DLA reported by \citet[filled circles]{coo10b}.
This DLA was originally identified in the SDSS spectrum
of the QSO J0035$-$0918. Follow-up high resolution spectroscopy indicated
 $z_{\rm abs}=2.3400972$, $\log N({\rm HI})$ $/{\rm cm}^{-2}=20.55 \pm 0.1$,
 and [Fe/H] $\simeq -3.04$.
1) A pronounced carbon enhancement [C/Fe]\,$\simeq +1.53$ is found, which is well reproduced with our faint SN/HN models.
Low-mass AGB stars ($1-4M_\odot$) can also provide such high [C/Fe].
However, such low-mass stars are unlikely to contribute at the redshift of the C-rich DLA.
With the star formation history in the solar neighborhood, [C/Fe] reaches the maximum value at $z=1.8$ due to the AGB contribution \citep{kob11agb}.
2) The [O/Fe] ratio is higher than for non-faint supernovae ([O/Fe] $\sim 0.5-0.6$ for $25M_\odot$) and is consistent with the faint SN model because of the smaller ejected Fe mass.
3) The [(Si,S)/Fe] ratios are similar to those of non-faint supernovae, and also consistent with the faint SN/HN models.
4) The low Al abundance strongly suggests that the enrichment source is not Pop II supernovae but primordial supernovae.
5) The low N abundance is also consistent with the faint SN/HN models, and the high [C/N] ratio cannot be explained with mass loss from rotating massive stars or intermediate-mass AGB stars ($\gtsim 4M_\odot$).

In order to discuss the detailed explosion mechanism of the supernova, it is necessary to obtain the elemental abundances of iron-peak elements.
For the C-rich DLA, because of the low metallicity, it is impossible to detect heavier elements than S except for Fe.
In Fig. \ref{fig:dla} we overplot the element abundances recently reported by \citet[open circles]{coo10} for the $z_{\rm abs}=1.62650$
DLA in front of the gravitationally lensed quasar UM637A.
This abundance pattern can also be explained with faint supernovae.

The dotted line is for the nucleosynthesis yields of the pair instability supernova (PISN) of a $170M_\odot$ star \citep{ume02}.
Compared with core-collapse supernovae, the abundance pattern of PISNe can be summarized as follow:
1) The odd-Z effect is much larger than $\sim 1$ dex.
2) ${\rm [(Si,S,Ar,Ca)/Fe]}$ are much larger than [(O, Mg)/Fe] because of more extensive explosive oxygen burning.
3) ${\rm [Cr/Fe]}$ is much larger because of the larger incomplete Si-burning region.
4) ${\rm [(Co,Zn)/Fe]}$ are much smaller because of the much larger ratio between the complete and incomplete Si-burning regions.
All of these characteristics disagree with the observed elemental abundances of these metal-poor DLAs.
Even at high-redshift, there is no signature of the existence of PISNe.
The [Si/C] for PISNe is as large as $+1.5$, which is also inconsistent with the observational estimate in the intergalactic medium (IGM, [Si/C] $\sim 0.77$).
The IGM abundance looks more consistent with normal (non-faint) core-collapse supernovae with [C/Fe] $\sim 0$ and [Si/Fe] $\sim 0.7$ \citep{kob06}.

In the early stages of chemical enrichment, the ISM is supposed to be highly inhomogeneous, so that the properties of the first objects can be directly extracted from the comparison between the observed elemental abundances and nucleosynthesis yields.
Since the DLA abundances reflect the chemical enrichment in gas-phase, the binary or accretion scenarios of the EMP stars do not work.
To explain the abundance pattern of the C-rich DLA, enrichment by primordial supernovae is the best solution.
It is interesting that the observed DLA abundance is very similar to those of EMP stars in the solar neighborhood including the ultra metal-poor star HE0557-4840 ([Fe/H] $=-4.75$, [C/Fe] $=+1.6$) and BD+44$^\circ$493.
Some of EMP stars in dSphs and the Galactic outer halo also show similar carbon enhancement at [Fe/H] $\ltsim -3$.
Chemical enrichment by the first stars in the first galaxies is likely to be driven by core-collapse supernovae from $\sim 20-50 M_\odot$ stars, which may be consistent with the latest simulations of primordial star formation \citep[e.g.,][]{gre11}.

\vspace*{-5mm}
\section{Fluorine Problem and Globular Clusters (GCs)}
\vspace*{-3mm}

Fluorine is an intriguing, though currently poorly studied element.
$^{19}$F is mainly produced by core and shell He-burning at $T\gtsim$ $1.5 \times 10^8$ K in low-mass and massive stars.
The production of F in AGB stars is 
highly mass dependent, where F production peaks at $\sim 3 M_\odot$ at
solar metallicity; in higher mass models F is
destroyed by $\alpha$-captures caused by the higher temperatures reached
during He-burning. 
In \citet{kob11f}, we have calculated the nucleosynthesis yields of core-collapse supernovae (SNe and HNe) including the $\nu$-process.
Most of the energy from core-collapse supernovae is released as neutrinos and anti-neutrinos ($\gtsim 10^{53}$ erg). 
The $\nu$-process does not affect the yields of major elements such as Fe and $\alpha$ elements, but it increases those of some elements such as B, F, K, Sc, V, Mn, and Ti.

\begin{figure}
\center
\includegraphics[width=6cm]{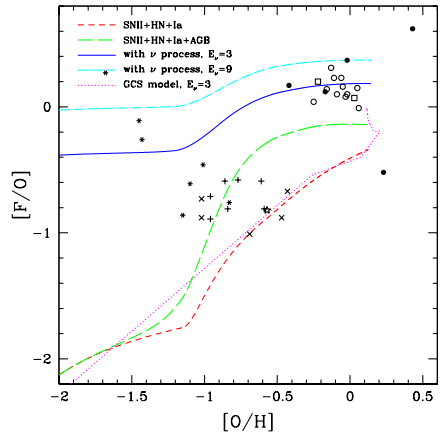}
\vspace*{-3mm}
\caption{\label{fig:fo}
Evolution of the [F/O] ratio against [O/H]
for the solar neighbourhood
with SNe II, HNe, and SNe Ia only (short-dashed lines), 
with AGB stars (long-dashed lines),
with the $\nu$-process of SNe II and HNe (solid line and dot-dashed line for $E_\nu=3 \times 10^{53}$ and $9 \times 10^{53}$ erg, respectively).
The dotted line is for the model for GCs.
See \citet{kob11f} for the observational data sources.
}
\vspace*{-5mm}
\end{figure}

Figure \ref{fig:fo} shows the evolution of [F/O] against [O/H].
Without the AGB yields and the $\nu$-process (short-dashed line), the predicted F abundance is too low to meet the observational data at all metallicities.
With the AGB yields (long-dashed line), [F/O] shows a rapid increase from [O/H] $\gtsim -1.2$ toward higher metallicities, which corresponds to the timescale of $2-4 M_\odot$ stars in the solar neighborhood.
At [O/H] $\sim 0$, [F/O] reaches $-0.14$, which is $0.26$ dex larger than the case without the AGB yields.
However, the present [F/O] ratio is still significantly lower than the observations at [O/H] $\sim 0$.
Note that 
the F yields from AGB stars were increased with updated reaction rates. 

The timescale of supernovae is much shorter than AGB stars, which means that the [F/O] ratio at low metallicities can be strongly enhanced by the $\nu$-process occurring in core-collapse supernovae.
With the standard neutrino luminosity $E_\nu=3 \times 10^{53}$ erg (solid line), the [F/O] ratio shows a plateau of [F/O] $\sim -0.4$ at [O/H] $\ltsim -1.2$, and reaches [F/O] $\sim +0.19$ at [O/H] $\gtsim 0$.
This is consistent with the observational data of field stars at $-0.5 \ltsim$ [O/H] $\ltsim 0$.
If we adopt a larger $E_\nu=9 \times 10^{53}$ erg (dot-dashed line), [F/O] can be as large as $\sim +0.37$ at [O/H] $\sim 0$.
In the bulge the star formation timescale is shorter and the average metallicity is higher than the solar neighborhood, but the [F/O] ratio is not so different at [O/H] $\sim 0$.
The observations for the bulge stars (filled circles) might suggest that the initial mass function is also different, although the number of observations is too small to make a conclusion.

At $-1 \ltsim$ [F/O] $\ltsim -0.5$ the observational data are for stars in GCs, 
which seem to be more consistent with the models with the AGB yields only than with the $\nu$-process.
However, it is unlikely that the existence of the $\nu$-process depends on the environment or metallicity.
The neutrino luminosity may be small in the case of faint supernovae with a large black hole, which give high [$\alpha$/Fe], but there is no significant difference seen in the  [$\alpha$/Fe] ratio between field halo stars and GC stars.
One possible scenario is as follows:
in GCs, the contribution from low-mass supernovae is smaller than in the filed.
Since the star formation occurs in a baryon dominated cloud with very high density, the initial star burst can be very intense.
After the initial star burst, because of the small gravitational potential, galactic winds are generated immediately after the explosion of massive supernovae, which may remove the contribution from low-mass supernovae.
The small production of $\alpha$ elements from low mass supernovae means that the [F/O] ratio can reach values as large as $\sim 0$.
In contrast, massive supernovae produce more $\alpha$ elements which results in [F/O] ratios of $\sim -0.5$, consistent with the observational data.
The dotted line shows an example of such a GC model (see \citealt{kob11f} for the details).

At low metallicities ([O/H] $\ltsim -1.2$) F is produced only from supernovae, and thus the observations of field stars at low-metallicities are important for constraining the neutrino luminosity released from a core-collapse supernova.
If the neutrino luminosity is specified, the F abundance along with C could be a good clock in the study of galactic archaeology to distinguish the contribution from AGB stars and supernovae.

\vspace*{-5mm}
\section{Dwarf Spheroidal Galaxies (dSphs)}
\vspace*{-3mm}

The F observations of stars in GCs suggest that the star formation and chemical enrichment histories of GCs are different from those of field stars and that low-mass supernova played a smaller role in shaping the chemical evolution of these systems.
This is the opposite to the situation for dSphs.
DSphs show low [$\alpha$/Fe] and low [Mn/Fe] \citep{rom11}, of which elemental abundance pattern is more consistent with the enrichment from low-mass SNe II than that from the SN Ia enrichment (\citealt{kob06}, see also Figure 20 of \citealt{kob11mw}).
There is no strong metallicity dependence on Mn yields of SNe Ia (T. Ohlubo, K. Nomoto et al, in prep.).
In dSphs, the dark matter component is large, the gas density is low, the star formation rate is low, and thus the contribution from massive supernovae is expected to be smaller than in the Milky Way halo.
The elemental abundance pattern (low [(Cu, Zn)/Fe]) of anomaly stars in the Galactic halo \citep{nis11} is also consistent with low-mass SNe II.

\end{document}